\documentclass[aps,prb,preprint,superscriptaddress,nobibnotes,amsmath,amssymb,amsfonts]{revtex4-1}

\usepackage{graphicx}
\usepackage{dcolumn}
\usepackage{bm}
\usepackage{multirow}
\usepackage{float}
\newcolumntype{C}[1]{>{\centering\arraybackslash}p{#1}}

\begin{document}

\title{Metastable Interlayer Frenkel Pair Defects by Dipole-like Strain Fields for Dimensional Distortion in Black Phosphorus}

\author{Devesh R. Kripalani}
\affiliation{School of Mechanical and Aerospace Engineering, Nanyang Technological University, Singapore 639798, Singapore}
\affiliation{Infineon Technologies Asia Pacific Pte Ltd, Singapore 349282, Singapore}

\author{Yongqing Cai}
\email[]{yongqingcai@um.edu.mo}
\affiliation{Joint Key Laboratory of the Ministry of Education, Institute of Applied Physics and Materials Engineering, University of Macau, Taipa, Macau, China}

\author{Ming Xue}
\affiliation{Infineon Technologies Asia Pacific Pte Ltd, Singapore 349282, Singapore}

\author{Kun Zhou}
\email[]{kzhou@ntu.edu.sg}
\affiliation{School of Mechanical and Aerospace Engineering, Nanyang Technological University, Singapore 639798, Singapore}


\begin{abstract}
The low formation energy of atomic vacancies in black phosphorus allows it to serve as an ideal prototypical system for exploring the dynamics of interlayer interstitial-vacancy (I-V) pairs (i.e. Frenkel defects) which account for Wigner energy release. Based on a few-layer model of black phosphorus, we conduct discrete geometry analysis and investigate the structural dynamics of intimate interlayer Frenkel pairs from first-principles calculations. We reveal a highly metastable I-V pair state driven by anisotropic dipole-like strain fields which can build strong connections between neighbouring layers. In the 2D limit (monolayer), the intimate I-V pair exhibits a relatively low formation energy of 1.54 eV and is energetically favoured over its isolated constituents by up to 1.68 eV. The barrier for annihilation of the Frenkel pair is 1.46 eV in the bilayer, which is remarkably higher than that of similar defects in graphite. The findings reported in this work suggest that there exist rich bridging pathways in black phosphorus, leading to stable dimensional reduction and structural condensation on exposure to moderate electron excitation or thermal annealing. This study paves the way for creating novel dimensional-hybrid polymorphs of phosphorus via the introduction of such metastable interlayer I-V pair defects.
\end{abstract}


\maketitle


\section{Introduction}
Frenkel defects, a type of coupled imperfection consisting of an interstitial atom and atomic vacancy (i.e. an I-V pair), are traditionally regarded as being rare in a three-dimensional (3D) lattice owing to the high energy cost for displacing an atom from its equilibrium position and lodging it within the tight confines of a nearby interstitial site.\cite{HKfrenkel98} Such intimate I-V pairs, mainly generated by particle irradiation, have long been predominantly found in ionic compounds having a large size difference between the anion and cation.\cite{KKfrenkel92} Seventy years ago, grave instances of neutron-induced damage of graphite in nuclear reactors triggered immense research into radiation defects in graphitic materials.\cite{Wnucl46,BNkogr56,Sirgr13} Irradiation causes dimensional changes and the accumulation of internal energy commonly known as the Wigner effect.\cite{Bkocnano99,TEivgr03} Spontaneous release of the Wigner energy in radiated graphite, as indicative of a release peak at $\sim$470 K, was later ascribed to the annihilation of Frenkel pair defects,\cite{MTgrwig65,ETivgr03} and is implicated in the Windscale reactor fire of 1957.\cite{Awindsc16} For two-dimensional (2D) layered materials, the van der Waals gap can facilitate the release of strain and accommodate the presence of interstitial atoms, leading to prolific I-V pairs as compared to their bulk 3D counterparts. Fortunately, under typical conditions of thermal and electric fields, the concentration of I-V pairs in graphite or multilayer graphene is generally low due to strong sp$^2$ carbon bonds, although experiments involving double-wall carbon nanotubes have since revealed their pronounced occurrence to be closely associated with lattice curvature.\cite{UScnt05}

Black phosphorus (or phosphorene in the case of a monolayer), another elemental layered non-ionic compound just like graphite, has recently emerged to great scientific acclaim as a promising semiconductor for next-generation optoelectronics owing to its band gap that remains direct for all thicknesses.\cite{CZtdbg14,LKphapp16,ZCphapp17} Remarkably, it features significantly low formation energy (\textit{E}\textsubscript{f} = 1.45 eV) and highly mobile character of vacancies (low activation barrier \textit{E}\textsubscript{a} = 0.30 eV) as compared to graphene (\textit{E}\textsubscript{f} = 7.57 eV, \textit{E}\textsubscript{a} = 1.39 eV).\cite{CKphos16} As the defect population is exponentially related to \textit{E}\textsubscript{f} according to the Boltzmann distribution, unlike its elemental cousin, graphite, black phosphorus tends to host Frenkel defects that may be solely introduced via thermal treatment, without the need for high-energy particle irradiation. The presence of curvature rooted in its inherent topography also suggests new, interesting opportunities of I-V pair defect engineering that may otherwise be hidden in other planar 2D materials. Moreover, under enhanced electric fields or electron injection near the proximity of contact, such defects are inevitably likely in real phosphorene-based devices.\cite{NLalox14,XQbpbeam17} Wigner energy-related structural relaxation of Frenkel pairs has also been indirectly referred to in previous studies of electrothermal-irradiated black phosphorus.\cite{LWbpthdp15,LLbbpthst17} While the Wigner effect and its adverse implications on performance could be largely avoided in graphene for normal nanoelectronic applications given the high formation energy of vacancies, the situation becomes much more critical for phosphorene due to its relatively low threshold against vacancy formation and migration. However, thus far, the structural characteristics of interlayer I-V pair defects, their kinetics, as well as energy release in black phosphorus are still unknown and have yet to be addressed at the atomic level.

\begin{figure*}
\includegraphics[width=0.9\textwidth]{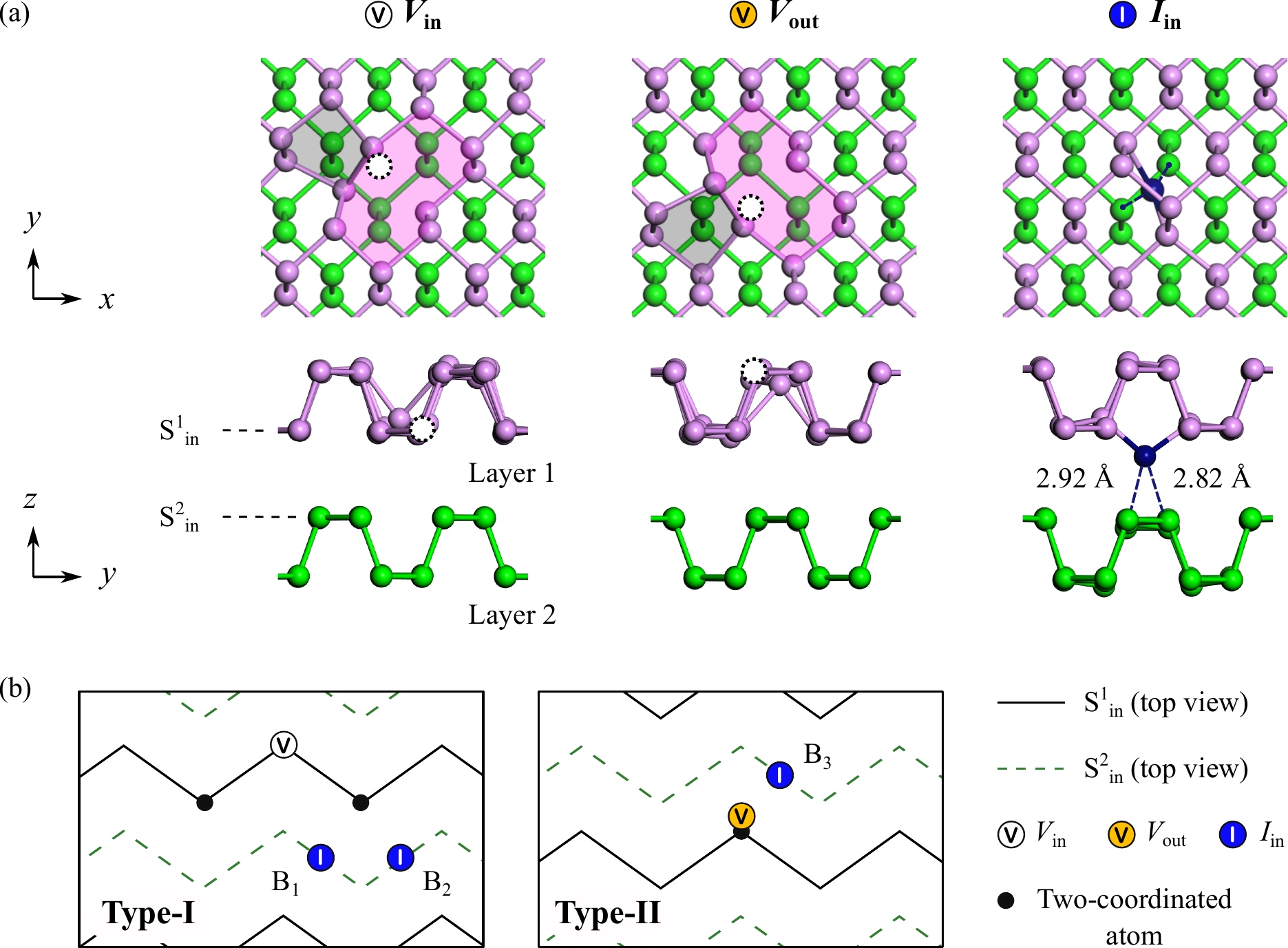}
\caption{(a) (From left to right) The relaxed structural configurations of bilayer phosphorene containing an inner vacancy \textit{V}\textsubscript{in}, outer vacancy \textit{V}\textsubscript{out} and inner self-interstitial \textit{I}\textsubscript{in}, respectively. Relative positions of the vacancy (white circles), pentagon-nonagon reconstructions (grey-pink shaded regions) and weak P-P bonds (blue dotted lines) are indicated accordingly. (b) Schematic diagrams of Type-I and Type-II I-V pair defects (top view), where only key structural details on the inner sub-lattices S$^{1}_{\textsubscript{in}}$ and S$^{2}_{\textsubscript{in}}$ are shown (see legend). \label{IVstructure}}
\end{figure*}

In this work, via first-principles calculations, we present theoretical evidence of metastable intimate interlayer I-V pair defects in few-layer phosphorene. Details of our computational method can be found in Section II of this article. By first examining the stacking registry of the bilayer (see Section I of the Supplemental Material (SM),\cite{SuppMater} and references\cite{DZbiph14,ZLbiphos15,RSuber83,WKpdef15} therein), the AB-stacked order is shown to bear the most energetically favourable structure. Based on the relative positions of point defects within the two layers, the following irreducible sites are identified, namely an inner vacancy \textit{V}\textsubscript{in}, outer vacancy \textit{V}\textsubscript{out} and inner self-interstitial \textit{I}\textsubscript{in}, as shown in Fig. \ref{IVstructure}(a). Current consensus have established that phosphorene, like graphene, adopts a pentagon-nonagon (5$\mid$9) reconstructed monovacancy (MV), which can remain stable at temperatures of up to 500 K.\cite{LWpdef05,HDpdef15,CKphos16} As for self-interstitials, they take up stable armchair-bridge positions between two host phosphorous atoms at the expense of one dangling bond.\cite{WKpdef15,VKpdef16} In the AB-stacked bilayer, the self-interstitial is additionally held in place by weak P-P bonds to its two nearest neighbours of the adjacent layer (i.e. layer 2). The point defects \textit{V}\textsubscript{in}, \textit{V}\textsubscript{out} and \textit{I}\textsubscript{in} are fundamental components of more complex I-V pairs and form the basis for which they can be investigated.

\section{Computational method}
Spin-polarized first-principles calculations are performed within the framework of density functional theory (DFT)\cite{KSexcorr65,PYdft94,SSdft09} using the Vienna \textit{ab initio} simulation package (VASP).\cite{KFvasp96} The optB88-vdW exchange-correlation functional\cite{DRvdwdf04,RSvdwdf09,KBvdwdf11,KBoptb8810} is adopted under the generalized gradient approximation (GGA) to account for the van der Waals forces between phosphorene layers. Our benchmark calculations on bulk black phosphorus have verified this approach, yielding lattice constants \textit{a} = 3.348 \AA, \textit{b} = 4.467 \AA\ and \textit{c} = 10.762 \AA\ which are in excellent agreement with experimental values (\textit{a} = 3.314 \AA, \textit{b} = 4.376 \AA, \textit{c} = 10.478 \AA).\citep{BRlatbp65} The optimized geometric parameters used in our slab models of phosphorene are given in Table S1 in the SM.\cite{SuppMater}

Vacancy and self-interstitial point defects are simulated in a 5 $\times$ 4 $\times$ 1 supercell. A kinetic energy cutoff of 500 eV is selected for the plane wave basis set, while the energy convergence criteria for electronic iterations is set at 10\textsuperscript{-6} eV. All structures are relaxed until the maximum Hellmann-Feynman force per atom is smaller than 0.02 eV/\AA. The first Bruillouin zone is sampled in reciprocal space with a 3 $\times$ 3 $\times$ 1 \textit{k}-point grid using the Monkhorst-Pack method. Periodic boundary conditions are applied in the zigzag (\textit{x}) and armchair (\textit{y}) in-plane directions, wheareas free boundary conditions are enforced in the normal (\textit{z}) direction by placing a vacuum layer of 15 \AA\ to eliminate spurious interactions between adjacent slabs. The climbing image nudged elastic band (NEB)\cite{HUneb00} method is employed to investigate the migration paths and energy barriers of I-V pair defects during recombination.

The defect formation energy \textit{E}\textsubscript{f} is given by Eq. (\ref{Ef}), where \textit{E}\textsubscript{p} and \textit{E}\textsubscript{d} refer to the total energy of the pristine and defective system respectively, while \textit{N}\textsubscript{P} indicates the number of phosphorous atoms added (\textit{N}\textsubscript{P} $>$ 0) or removed (\textit{N}\textsubscript{P} $<$ 0) to create the defect. Here, the chemical potential of phosphorous $\mu\textsubscript{P}$ is taken to be its atomic energy in the pristine system, \textit{q} is the charge state of the defect, and \textit{E}\textsubscript{Fermi} is the Fermi level or electron chemical potential representing the energy of the reservoir for electron exchange. Note that \textit{E}\textsubscript{Fermi} is referenced to the energy level of the valence band maximum $\epsilon$\textsubscript{VBM} and \textit{$\Delta V$} accounts for corrections due to the image-charge and potential alignment of the charged defect. Finite-size corrections are adopted by using the method of Makov and Payne.\cite{MPsizecorr95}

{\small
\begin{align}
\textit{E}\textsubscript{f} = \textit{E}\textsubscript{d} - (\textit{E}\textsubscript{p}+\mu\textsubscript{P}\textit{N}\textsubscript{P}) + \textit{q}(\textit{E}\textsubscript{Fermi} + \epsilon\textsubscript{VBM} + \textit{$\Delta V$}) \label{Ef}
\end{align}
}

The interaction energy \textit{E}\textsubscript{i} reflects the energy cost of dissociating an I-V pair into its constituent point defects (i.e. an infinitely-separated I-V pair). This quantity can be calculated by Eq. (\ref{Ei}), where \textit{E}\textsubscript{I}, \textit{E}\textsubscript{V} and \textit{E}\textsubscript{I-V} denote the total energy of the defective system containing a self-interstitial, vacancy and an I-V pair, respectively.

{\small
\begin{align}
\textit{E}\textsubscript{i} = (\textit{E}\textsubscript{I}+\textit{E}\textsubscript{V})-(\textit{E}\textsubscript{I-V}+\textit{E}\textsubscript{p}) \label{Ei}
\end{align}
}

\begin{figure*}
\includegraphics[width=0.9\textwidth]{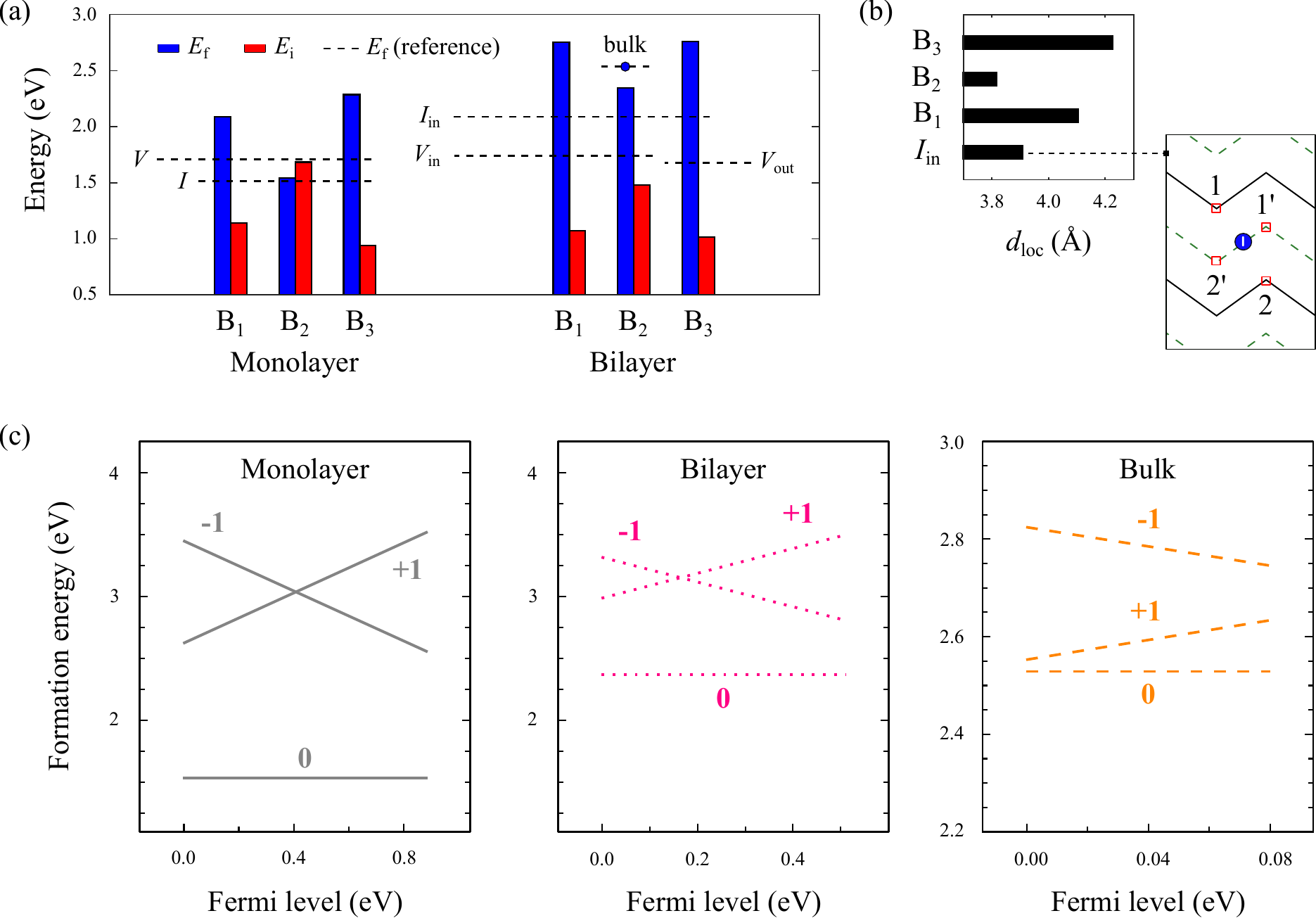}
\caption{(a) The energetics, and (b) local interlayer distance at the interstitial site \textit{d}\textsubscript{loc} of Frenkel defects considered in this work. The schematic in (b) indicates the nearest neighbours to the interstitial site for the reference case \textit{I}\textsubscript{in}. (c) Formation energy of neutral (0) and charged ($\pm$1) B$_2$ I-V pair defects in monolayer (grey solid lines), bilayer (purple dotted lines) and bulk (orange dashed lines) phosphorene. The Fermi level ranges across the band gap from the valence band maximum (VBM) to the conduction band minimum (CBM). \label{IVenergetics}}
\end{figure*}

\section{Results and Discussion}
\subsection{Frenkel defects in black phosphorus: Structural variants and energetics}
Herein, our focus lies on close-proximity I-V configurations, in which vacancy-induced two-coordinated atoms strongly interact with vicinal self-interstitials. Particularly, we can classify I-V pair defects of this nature into two broad categories: Type-I and Type-II, depending on the relative location of the vacancy with respect to the self-interstitial. The configurational details (top view) of bilayer phosphorene pertaining to each of these categories are schematically illustrated in Fig. \ref{IVstructure}(b), with only key topographic features depicted for visual clarity. As shown, Type-I I-V pairs consist of \textit{V}\textsubscript{in} and \textit{I}\textsubscript{in} point defects, and present two symmetry-inequivalent armchair-bridge sites, B$_1$ and B$_2$, for self-interstitials. Conversely, Type-II I-V pairs entail \textit{V}\textsubscript{out} and \textit{I}\textsubscript{in} point defects, and host one unique bonding site, B$_3$, for self-interstitials. All three defect configurations B$_1$, B$_2$ and B$_3$ are determined to be metastable at 0 K, showing no signs of spontaneous decomposition or recombination into its pristine state. Their energetics in both the monolayer and bilayer are presented in Fig. \ref{IVenergetics}(a).

The formation energy of I-V pairs is found to be minimized in the B$_2$ configuration and can be as low as 1.54 eV on the free surface of the monolayer. While a recent theoretical study by Gaberle and Shluger has contemplated the prospect of Frenkel defects in phosphorene, substantially higher \textit{E}\textsubscript{f} of $\sim$3 eV had been predicted,\cite{GSfrenkel18} possibly without fully accounting for interdefect coupling. Furthermore, our results indicate that the coalescence between self-interstitials and vacancies is energetically favoured in the monolayer to form B$_2$ I-V pairs. However, in the bilayer and beyond, intimate I-V pairs tend to possess much higher \textit{E}\textsubscript{f} than their fundamental point defects. This implies a higher concentration of isolated defects than that of Frenkel pairs under equilibrium conditions, which explains the general lack of evidence of Frenkel defects in practical samples of black phosphorus following standard growth processes.\cite{KHstmbp17} Instead, Frenkel defects could be generated under high-temperature non-equilibrium conditions, such as during electron irradiation whereby the lattice is exposed to significant knock-on bombardment events,\cite{VKpdef16} or via high-temperature annealing in vacuum.\cite{LWbpthdp15} Appropriate thermal annealing would also help promote the stable formation of Frenkel pairs due to the exothermic process of coupling between isolated defects, as reflected by their positive interaction energies \textit{E}\textsubscript{i} (see Fig. \ref{IVenergetics}(a)).

The interaction energy is at its highest for the B$_2$ structure, with values of up to 1.68 eV and 1.48 eV in the monolayer and bilayer, respectively. For a deeper insight into the differences between B$_1$, B$_2$ and B$_3$ I-V pairs in the bilayer, we examine the local interlayer distance at the interstitial site \textit{d}\textsubscript{loc}, as shown in Fig. \ref{IVenergetics}(b). This geometric parameter is defined by \textit{d}\textsubscript{loc} = (\small$(\textit{z}_{1}+\textit{z}_{2})-(\textit{z}_{1'}+\textit{z}_{2'})$\normalsize)/2, where \textit{z}$_i$ is the \textit{z}-coordinate of the neighbouring atom \textit{i} to the self-interstitial (see given schematic in Fig. \ref{IVenergetics}(b) for the reference case \textit{I}\textsubscript{in}). The values of \textit{d}\textsubscript{loc} for B$_1$ (4.10 \AA) and B$_3$ (4.23 \AA) are observed to be much larger than that of \textit{I}\textsubscript{in} (3.91 \AA), while that for B$_2$ (3.82 \AA) is relatively small in comparison. This suggests that the stabilising effects of interlayer binding are the strongest in the B$_2$ structure such that it displays the highest resistance against dissociative and recombinative mechanisms, especially under elevated temperatures. Hence, in agreement with earlier analysis, the B$_2$ I-V pair is the most energetically preferred adaptation of the Frenkel defect. Moreover, by performing finite temperature (canonical ensemble) \textit{ab initio} molecular dynamics simulations at 300 K and for a period of 10 ps, we establish that the B$_2$ I-V pair can remain thermodynamically stable at room temperature (see Section II of the SM,\cite{SuppMater} and references\cite{Tmdrmsd05,Wmdrmsf62} therein).

In Fig. \ref{IVenergetics}(c), we consider the effect of net charge on the formation energy of the B$_2$ I-V pair in monolayer, bilayer and bulk phosphorene. For all cases, both negatively ($-$1) and positively ($+$1) charged states are energetically unstable as compared to the defect in its neutral (0) condition. This implies that, in the near proximity for forming a Frenkel pair, deviation of the phosphorus atom from its equilibrium site tends not to alter the electrostatic environment of the host lattice. Such neutral tendency is reasonable for a non-polar homoelemental sheet like phosphorene as the Frenkel pair defect can remain pseudo self-compensated with weak charge accumulation/donation.

\subsection{Electronic properties of Frenkel defects in bilayer phosphorene}
\begin{figure*}
\includegraphics[width=0.9\textwidth]{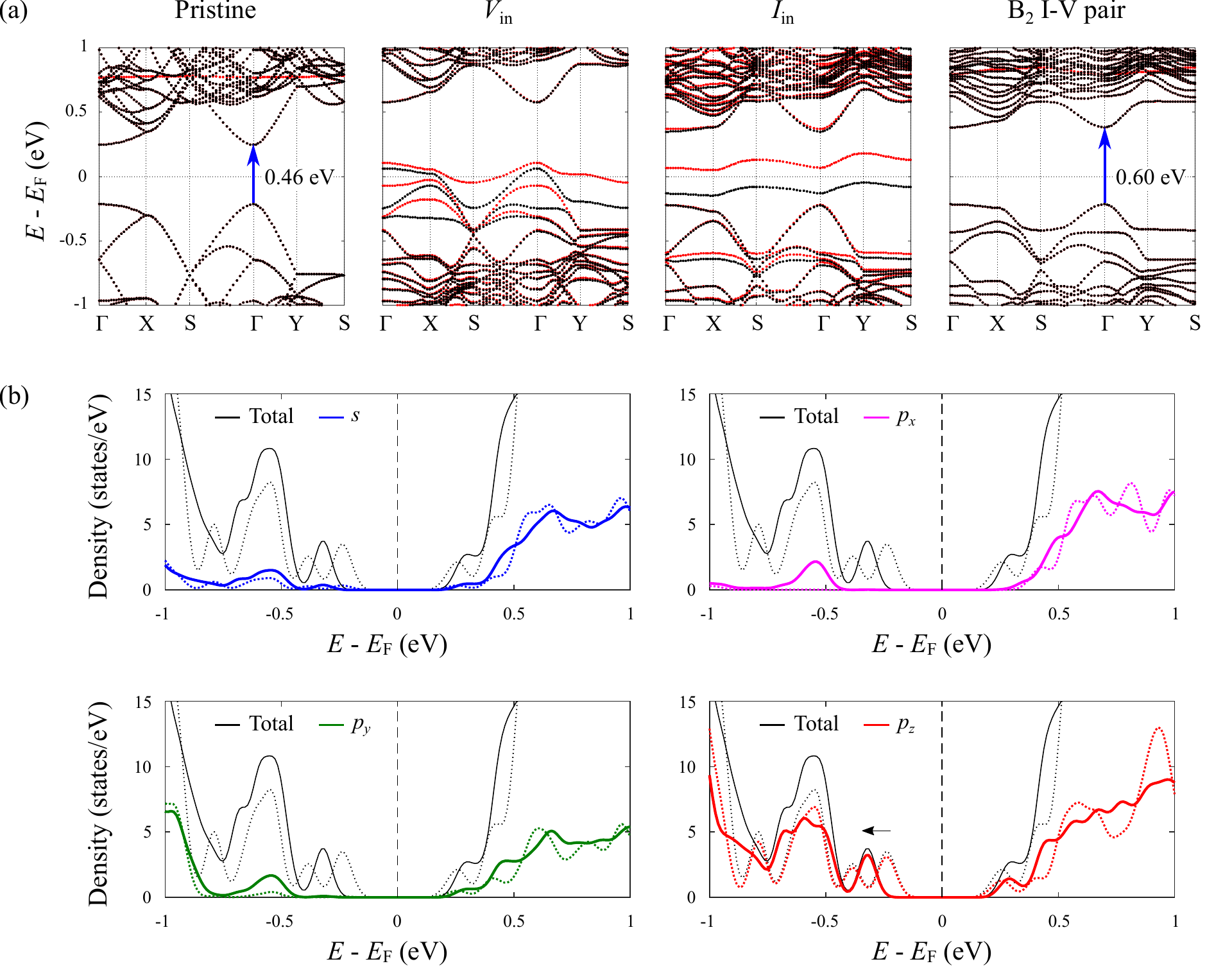}
\caption{(a) (From left to right) Spin-polarized electronic band structures of bilayer phosphorene containing no defects (pristine), an inner vacancy \textit{V}\textsubscript{in}, inner self-interstitial \textit{I}\textsubscript{in} and a B$_2$ I-V pair, respectively. Spin-up (red) and spin-down (black) band components are plotted relative to the Fermi level \textit{E}\textsubscript{F}. (b) The total and orbital projected density of states of bilayer phosphorene with (solid line) and without (dotted line) the B$_2$ I-V pair defect. The dashed line indicates the Fermi level \textit{E}\textsubscript{F} corresponding to the pristine bilayer. \label{Bands}}
\end{figure*}

In this section, the effect of interdefect coupling on the electronic band structure of bilayer phosphorene is investigated. Spin-polarized band structures are calculated in Fig. \ref{Bands}(a) along the high symmetry path $\Gamma$-X-S-$\Gamma$-Y-S defined for a 2D orthorhombic lattice. Note that while the optB88-vdW functional tends to underestimate the band gap (see Table S2 in the SM,\cite{SuppMater} and references\cite{CZtdbg14,DZbiph14,PBE96,HSE03,HSE06,QKbgph14,LNmob14,Kbgexpph53,Wbgexpph63,MSbgexpph81,NAbgexpph83} therein), it remains sufficiently reliable for making qualitative predictions for the purposes of our study.\cite{CZtdbg14,PDsbph17} In its pristine state, bilayer phosphorene is a non-magnetic semiconductor with a direct band gap (0.46 eV) at the $\Gamma$ point. The presence of an inner vacancy (\textit{V}\textsubscript{in}) is shown to introduce defect states that intersect the Fermi level, reflective of a metallic system. On the other hand, for the case of an inner self-interstitial (\textit{I}\textsubscript{in}), the band structure contains two defect bands which straddle the Fermi level. Both \textit{V}\textsubscript{in} and \textit{I}\textsubscript{in} are spin-polarized with net magnetic moments of 0.39 $\mu$\textsubscript{B} and 0.97 $\mu$\textsubscript{B}, respectively. Interestingly, interdefect coupling in the form of the B$_{2}$ I-V pair is found to effectively quench the net magnetic moments originating from the fundamental point defects, while widening the direct band gap at the $\Gamma$ point to 0.60 eV (blue-shift).

The underlying physical mechanism behind the widening of the band gap is ascribed to the breaking of lattice periodicity in the vicinity of the Frenkel pair, through (i) a (5$\mid$9) reconstruction at the vacancy site, and (ii) a localized increase in the interlayer distance of the bilayer structure such that it essentially behaves as two decoupled monolayers in the defect neighbourhood. As discussed previously (see Fig. \ref{IVenergetics}(b)), the local interlayer distance associated with the B$_{2}$ I-V pair can be as high as 3.82 \AA\ at the self-interstitial site, significantly exceeding the equilibrium separation distance of 3.20 \AA\ in the pristine bilayer. This mechanism is further corroborated by an analysis of the projected density of states, as shown in Fig. \ref{Bands}(b). Notably, we see that the regions surrounding the valence band maximum and conduction band minimum are primarily dominated by \textit{p}\textsubscript{\textit{z}} orbital states (red), which illustrate the strong role of interlayer interactions on the electronic properties of black phosphorus. In this regard, the widening of the band gap is attributed to a net shift of the valence band edge towards lower energy levels (black arrow), while that of the conduction band remains relatively fixed with respect to the Fermi level.

\begin{figure*}
\includegraphics[width=0.6\textwidth]{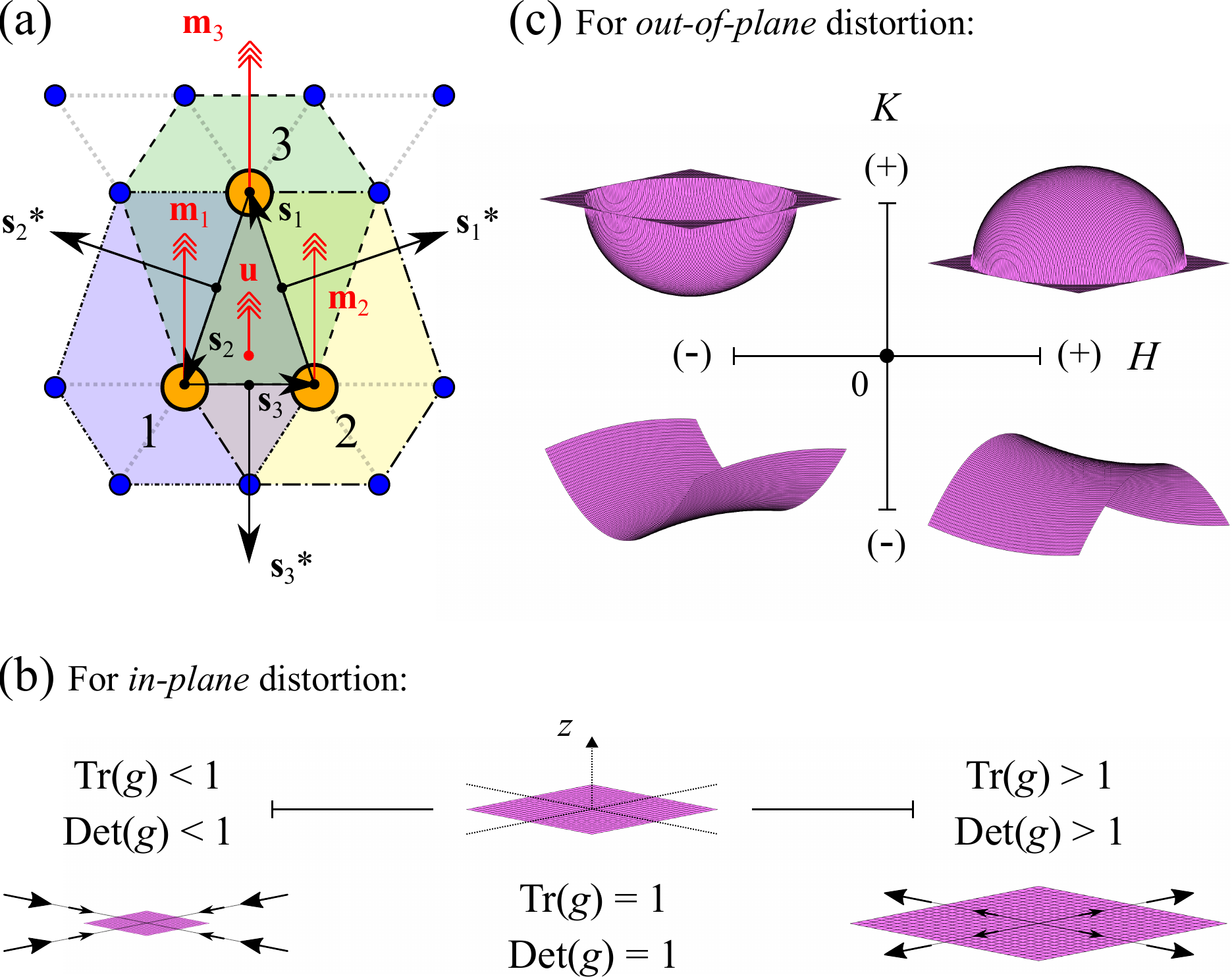}
\caption{(a) Discrete geometry formulation for a 2D lattice. Each triangulated element is defined by the vertex atoms 1, 2 and 3 (orange) and their nearest neighbours (blue). (b)-(c) Characteristic surface profiles associated with the four invariants Tr(\textit{g}), Det(\textit{g}), \textit{H} and \textit{K}. \label{DGA_scheme}}
\end{figure*}

\begin{figure*}
\includegraphics[width=0.9\textwidth]{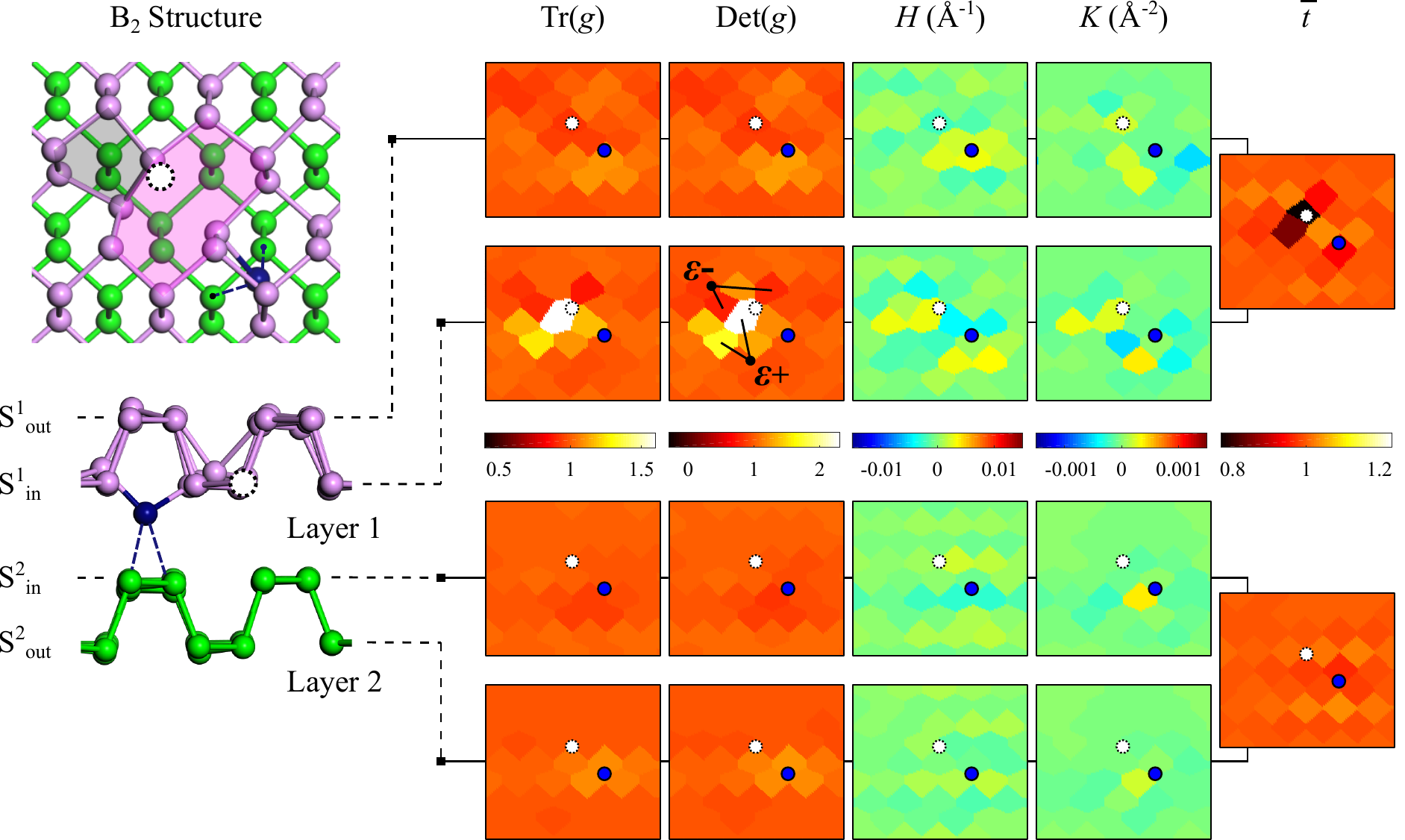}
\caption{The relaxed structural configuration of the B$_2$ I-V pair and an analysis of its discrete geometry (top view). Relative positions of the vacancy and self-interstitial on the surface are indicated by white and blue circles, respectively. \label{DGA_results}}
\end{figure*}

\subsection{Topographic analysis of Frenkel defects based on a discretization approach}
The local atomistic structure of the B$_{2}$ I-V pair is analysed using discrete geometry within the context of 2D material nets.\cite{Ldg2d97,BSdg2d08} As opposed to conventional continuum-based methods for parameterizing the strain field, discrete geometry goes beyond first order elasticity to provide an exact description of shape under arbitrary deformations. The discrete geometry of a surface is characterized by four invariants of its metric tensor \textit{g} and curvature tensor \textit{k}, denoted by the trace and determinant of \textit{g} (i.e. Tr(\textit{g}) and Det(\textit{g})), mean curvature \textit{H} and Gaussian curvature \textit{K}.

The discrete geometry is formulated using the method of triangulations over a finite mesh of atomic positions,\cite{BSdg2d08,WTdga12,MUdginv15} as illustrated in Fig. \ref{DGA_scheme}(a). Each triangulated element comprises of three adjacent atoms (denoted by atom index \textit{p} = 1, 2, 3), and is defined by the directed edges \textbf{s}$_1$, \textbf{s}$_2$ and \textbf{s}$_3$ such that $\textbf{s}_1 + \textbf{s}_2 + \textbf{s}_3 = 0$. A discrete analogue of the infinitesimal length \textit{ds}$^2$ is given by ${\textit{Z}_{p}}^\text{I} = \textbf{s}_p \cdot \textbf{s}_p$, which yields the square of the shortest distance between atoms. The variation in orientation between the normal vectors $\textbf{m}_{q}$ and $\textbf{m}_{r}$ is projected onto their common edge \textbf{s}$_p$ following the relation ${\textit{Z}_{p}}^\text{II} = (\textbf{m}_{r} - \textbf{m}_{q}) \cdot \textbf{s}_p$. Here, $\textbf{m}_{p}$ is the mean of the normal vectors of triangulated elements which immediately surround atom \textit{p} (see dashed outlines). Another relevant quantity, known as the dual edge, is given by $\textbf{s}_{p}^{*} = \textbf{s}_{p} \times \textbf{u}$, where $\textbf{u}$ is the normal vector of the triangulation described by atoms 1, 2 and 3. The area of the triangulated element in its pristine (reference) condition is denoted by \textit{A}$_0$, while that under arbitrary deformation is indicated by \textit{A}$_1$. The metric tensor \textit{g} and curvature tensor \textit{k} can then be expressed as follows:

{\small
\begin{align}\label{DG_gk}
\textit{g} & = -\frac{1}{8{\textit{A}_{0}}^{2}}\sum_{(p,q,r)}({\textit{Z}_{p}}^\text{I}-{\textit{Z}_{q}}^\text{I}-{\textit{Z}_{r}}^\text{I}) \ \textbf{s}_{p}^{*}\otimes\textbf{s}_{p}^{*} \\
\textit{k} & = -\frac{1}{8{\textit{A}_{1}}^{2}}\sum_{(p,q,r)}({\textit{Z}_{p}}^\text{II}-{\textit{Z}_{q}}^\text{II}-{\textit{Z}_{r}}^\text{II}) \ \textbf{s}_{p}^{*}\otimes\textbf{s}_{p}^{*}
\end{align}
}

The resulting tensors \textit{g} and \textit{k} are in the form of 3 $\times$ 3 matrices, with the parentheses (\textit{p}, \textit{q}, \textit{r}) representing the set of three contributions (1, 2, 3), (2, 3, 1) and (3, 1, 2) included in the summation. The eigenvalues \{\textit{k}$_1$, \textit{k}$_2$, 0\} of the curvature tensor \textit{k} provide the principal curvatures \textit{k}$_1$ and \textit{k}$_2$ for each triangulation such that $\textit{H} = (\textit{k}_1 + \textit{k}_2) / 2$ and $\textit{K} = \textit{k}_1 \textit{k}_2$ can be evaluated. The four invariants Tr(\textit{g}), Det(\textit{g}), \textit{H} and \textit{K} corresponding to each atomic position are averaged over their respective values at all triangulated elements that share the same vertex. As shown in Figs. \ref{DGA_scheme}(b)-(c), the reference values of these invariants are Tr(\textit{g}) = 1, Det(\textit{g}) = 1, \textit{H} = 0 and \textit{K} = 0, which correspond to the case of pristine phosphorene in its strain-free, planar configuration. Tr(\textit{g}) and Det(\textit{g}) take on values greater (less) than 1 for varying degrees of in-plane tension (compression), while \textit{H} and \textit{K}, in the usual mathematical sense, measure the out-of-plane deviation from planarity and character of the surface profile (i.e. elliptical (\textit{K} $>$ 0) or hyperbolic (\textit{K} $<$ 0)). Due to the puckered structure of phosphorene, the normalized layer thickness \textit{$ \overline{t}$} provides another quantity of interest for our analysis.

The discrete geometry of the B$_{2}$ I-V pair is evaluated for the inner and outer sub-lattice of each layer individually, as shown in Fig. \ref{DGA_results}. Note that these results correspond to the case of the defect in the dilute limit and are independent of supercell size (see Fig. S3 in the SM).\cite{SuppMater} From Tr(\textit{g}) and Det(\textit{g}), we find that the vacancy-induced (5$\mid$9) reconstruction generates significant tensile strain in S$^{1}{\textsubscript{in}}$, while the encapsulation of the self-interstitial yields low-to-moderate compression (tension) in the inner (outer) sub-lattice of both layers. The resulting strain distribution consisting of local tensile ($\epsilon+$) and compressive ($\epsilon-$) zones creates a dipole-like mechanism which drives the coupling and binding of the Frenkel pair. Since the strain fields imposed by the vacancy and self-interstitial are highly anisotropic, the coupling strength of Frenkel defects will strongly depend on the relative arrangement of interstitials and vacancy sites in the lattice.

Another plausible factor that accounts for the binding of Frenkel defects is related to the electronic interaction. It is noted that both isolated vacancies and self-interstitials are accompanied with spin-polarized dangling states (see Fig. \ref{Bands}(a)). However, upon formation of the Frenkel pair, such shallow empty defect states are saturated and a new hybridized state is formed and buried within the valence band; accordingly, the I-V pair is stabilized. In addition, the coupling between the vacancy and self-interstitial gives rise to regions of strong, I-V oriented rippling in both the inner and outer sub-lattices of layer 1, as reflected by \textit{H} and \textit{K}. The structural signature of the B$_{2}$ I-V pair is also shaped by changes to the layer thickness, which decreases by up to 20\% in the neighbourhood of the vacancy in layer 1. The distinct out-of-plane dimensional distortion introduced by such defects can enhance the chemical activity of locally rippled regions around vacancy sites.\cite{URenv15,KCphos16} Interestingly, this may serve to facilitate the substitutional doping process during ion implantation treatments of phosphorene-based semiconductors.

\begin{figure*}
\includegraphics[width=0.9\textwidth]{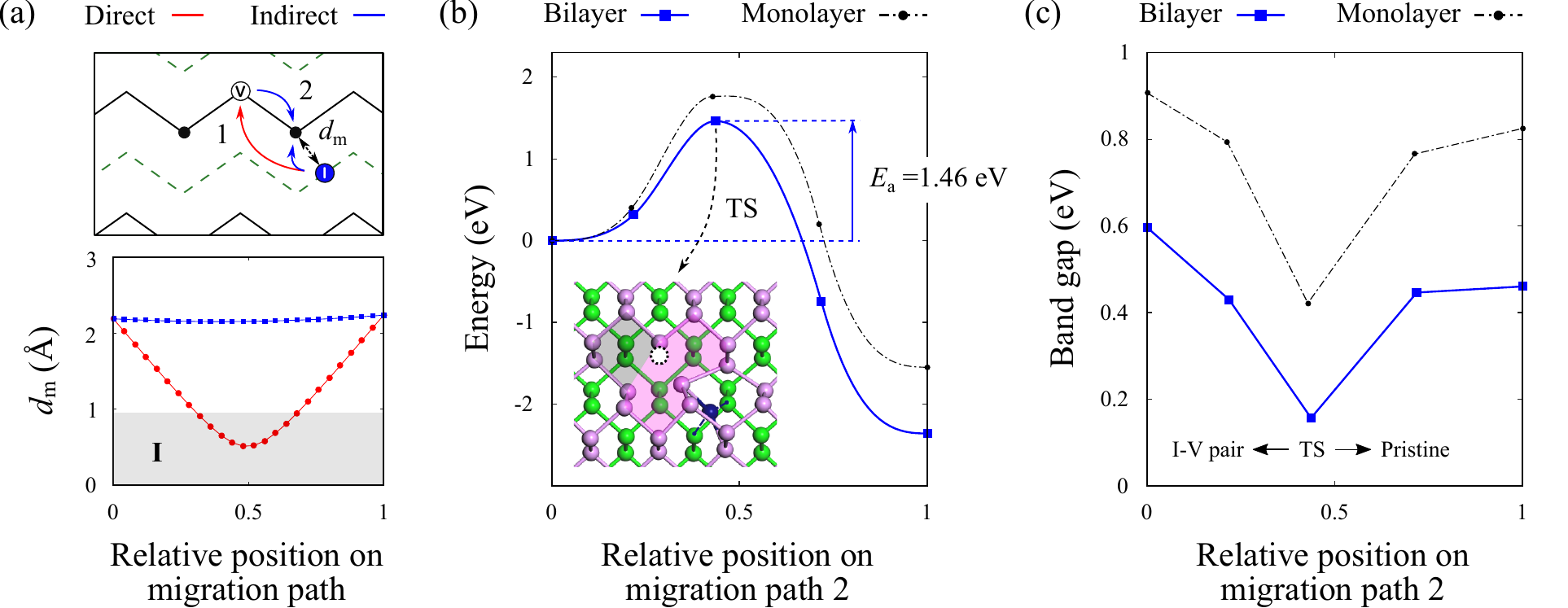}
\caption{(a) Top panel: Schematic diagram showing the direct (path 1) (red) and indirect (path 2) (blue) routes to recombination of the B$_2$ I-V pair. Bottom panel: The variation in bond length \textit{d}\textsubscript{m} between the self-interstitial and two-coordinated host atom during migration. Region I (shaded grey) denotes the range of \textit{d}\textsubscript{m} less than the atomic radius of phosphorus. (b) Energy profiles, and (c) band gap evolution of the indirect path to recombination, with the transition state (TS) given in the inset of (b). The solid (blue) and dash-dotted (black) lines correspond to the case of the bilayer and monolayer, respectively. \label{BRcom}}
\end{figure*}

\subsection{Barriers to recombination of Frenkel defects in phosphorene}
By virtue of the close proximity between vacancies and self-interstitials in the B$_{2}$ I-V pair, short-range recombination can take place. In phosphorene, recombination events may also be driven by the inherent presence of highly mobile vacancies, which are particularly faster along the zigzag direction.\cite{CKphos16} Here, two paths to recombination - direct (red) and indirect (blue) - are preliminarily proposed, as indicated in Fig. \ref{BRcom}(a) (top panel). In the direct route, the migrating self-interstitial simply recombines at the inner vacancy site of layer 1. On the other hand, in the indirect route, both the vacancy and self-interstitial undergo simultaneous migration towards one another such that recombination takes place at the original site of the two-coordinated host atom.

The variation in bond length \textit{d}\textsubscript{m} between the self-interstitial and two-coordinated host atom during migration is monitored in Fig. \ref{BRcom}(a) (bottom panel). Region I (shaded grey) denotes the range of \textit{d}\textsubscript{m} which is less than the atomic radius of P (0.95 \AA), and is therefore highly unphysical. Clearly, the direct route appears to be energetically unfavourable owing to the excessive overlap of atomic cores for a large part of the recombination process. In contrast, the indirect route presents a more viable mechanism as the bond length \textit{d}\textsubscript{m} is well-preserved throughout the course of migration. The energy profiles corresponding to the indirect path to recombination are shown in Fig. \ref{BRcom}(b), while the transition state (TS) is given in the inset. We determine the activation energy \textit{E}\textsubscript{a} to be 1.46 eV in the recovery of the pristine bilayer, with a net energy release of $\sim$2.4 eV. By thinning down phosphorene even further into its monolayer, the recombination barrier increases to 1.77 eV due to quantum size effects that enhance the strength of P-P bonding. The recombination barrier of phosphorene-based I-V pairs is found to be higher than that reported for atomically-flat graphite (\textit{E}\textsubscript{a} = 1.30 eV) which is responsible for Wigner energy release at 470 K.\cite{MTgrwig65,ETivgr03} This is surprising considering that graphitic C-C bonds are generally much stronger than that of P-P in phosphorene, and goes to show the pivotal role of lattice curvature in the formation of highly metastable Frenkel defects.

Our results are consistent with recent reports on the formation of skeletal structures of black phosphorus annealed at around 670 K.\cite{LWbpthdp15} The nature of such dimensional distortion has long been open to speculation as an amorphous red phosphorus-like skeleton. Here, we suggest that this structure could be closely linked with the onset and development of intimate interlayer I-V pair defects. While not been studied, these complex defects could even lead to higher-order defect rearrangements and spontaneous release of energy with either electron irradiation or thermal treatment.

On closer inspection of the intermediate states during recombination, we are able to obtain a clearer understanding of how the band gap might evolve with structure (see Fig. \ref{BRcom}(c)). Lattice instabilities and distortion, which tend to introduce defect states near the Fermi level, ultimately lead to minimum band gap values at the TS for both the monolayer (0.42 eV) and bilayer (0.16 eV). The paths connecting the TS to either the I-V pair reconstruction or pristine lattice seem to provide alternative modes by which atomic orbitals can interact and hybridize, thereby giving rise to entirely different band gap widths at equilibrium in each case. Through verification of the band structure, we can also attest that in low concentrations, the presence of Frenkel defects should not alter the direct-gap character of phosphorene at the $\Gamma$ point.

\section{Conclusion}
Overall, this work provides fresh insights into possible metastable I-V pair defects in black phosphorus. As opposed to graphene for which Frenkel defects are unlikely to proliferate and hamper normal nanoelectronic applications, phosphorene-based devices are prone to such lattice imperfections and may suffer from consequent defect rearrangements and structural condensation on exposure to thermal and electric fields. It is also likely that such metastable I-V pairs may form the crux of defect clusters which can extend in both in-plane and out-of-plane directions, particularly due to prolonged irradiation and at high temperatures.\cite{TDextiv13} The generation of I-V pairs is also a desirable way for producing novel polymorphs of mixed dimensional character, opening up new, non-conventional avenues for defect engineering applications. The findings reported here illuminate the structural complexities of strongly-coupled intrinsic defects in black phosphorus, and cater to growing experimental interest to integrate phosphorene into high-performance optoelectronics.

\begin{acknowledgments}
This work was supported by the Economic Development Board, Singapore and Infineon Technologies Asia Pacific Pte Ltd through the Industrial Postgraduate Programme with Nanyang Technological University, Singapore and the Ministry of Education, Singapore (Academic Research Fund TIER 1 - RG174/15). The computational calculations for this work was partially performed on resources of the National Supercomputing Centre, Singapore (https://www.nscc.sg).
\end{acknowledgments}

\bibliography{bib_manuscript}

\end{document}